\documentclass[12pt]{article}

\usepackage{scicite}

\usepackage{times}
\usepackage{url}
\usepackage{graphicx} 
\usepackage{subfigure}
\usepackage{array}
\usepackage{longtable}

\newenvironment{sciabstract}{%
\begin{quote} \bf}
{\end{quote}}

\newcommand{\fig}[4]{
    \begin{figure}[tb]\centering
    \includegraphics[width=#4in]{fig/#1}
    \vspace{0ex}
    \caption{#2}
    \label{#3}\end{figure}
    {}}

\newcommand{\figtwo}[8]{
    \begin{figure}[tb]\centering
    \subfigure[#2] {
    \label{#3}
    \includegraphics[width=1.55in]{fig/#1}
    }
    \subfigure[#5] {
    \label{#6}
    \includegraphics[width=1.55in]{fig/#4}
    }
    \vspace{0ex}
    \caption{#7}
    \label{#8}
    \end{figure}
    {}}

\newcounter{lastnote}

\title{ExpertSeer: a Keyphrase Based Expert Recommender for Digital Libraries}

\author
{Hung-Hsuan Chen,$^{1}$ Alexander G. Ororbia II,$^{2}$ C.\ Lee Giles$^{2}$\\
\\
\normalsize{$^{1}$Computational Intelligence Technology Center,}\\
\normalsize{Industrial Technology Research Institute, Taiwan}\\
\normalsize{$^{2}$Information Sciences and Technology,}\\
\normalsize{Pennsylvania State University, University Park, Pennsylvania, USA}\\
\normalsize{\url{hhchen1105@gmail.com}, \url{ago109@ist.psu.edu}, \url{giles@ist.psu.edu}}
}

\date{}

\begin{document}


\baselineskip24pt


\maketitle


\begin{sciabstract}
\textmd{
We describe ExpertSeer, a generic framework for expert recommendation
based on the contents of a digital library.  Given a query term $q$, ExpertSeer recommends
experts of $q$ by retrieving authors who published relevant papers determined
by related keyphrases and the quality of papers.  The system is based on a
simple yet effective keyphrase extractor and the Bayes' rule for expert
recommendation.  ExpertSeer is domain independent and can be applied to different
disciplines and applications since the system is automated and not tailored to a
specific discipline.  Digital library providers can employ the system to enrich
their services and organizations can discover experts of interest within
an organization.  To demonstrate the power of ExpertSeer, we apply the
framework to build two expert recommender systems.  The first, CSSeer, utilizes
the CiteSeerX digital library to recommend experts primarily in computer science.
The second, ChemSeer, uses publicly available documents from the Royal Society
of Chemistry (RSC) to recommend experts in chemistry.  Using one thousand computer
science terms as benchmark queries, we compared the top-$n$ experts ($n=3, 5, 10$)
returned by CSSeer to two other expert recommenders -- Microsoft Academic Search
and ArnetMiner -- and a simulator that imitates the ranking function of Google Scholar.
Although CSSeer, Microsoft Academic Search, and ArnetMiner mostly return prestigious
researchers who published several papers related to the query term, it was found that
different expert recommenders return moderately different recommendations.  To further
study their performance, we obtained a widely used benchmark dataset as the
ground truth for comparison.  The results show that our system outperforms
Microsoft Academic Search and ArnetMiner in terms of Precision-at-$k$ ($P@k$)
for $k=3, 5, 10$.  We also conducted several case studies to validate the
usefulness of our system.
}
\end{sciabstract}


\section{Introduction}

Business organizations depend heavily on information technology to
analyze data, manage resources and perform knowledge discovery~\cite{ullah2013systematic} \cite{lim2013business}.  Studies have
shown that companies can improve their market performance by appropriately
managing their knowledge and digital property~\cite{wu2013does}.  Here,
we propose a framework to manage an important resource of organizations --
an organization's experts -- based on documents or technical reports of an organization.
As such, companies can better optimize personnel utilization through this
framework.  Finding experts is also important in academia when assistance in
answering difficult questions is required, members for a committee (such as for
a conference) need to be found, or there is simply an interest in identifying
experts of a given domain for knowledge discovery.

Early expert recommender systems depended on manually constructed databases
that stored the skills of individuals.  However, manual methods do not easily scale.  In addition, the list could be biased and limited by the
compiler's knowledge of the domain topic.  As a result, recent research has
focused on automated expert finding~\cite{balog2006formal},
\cite{deng2008formal}, \cite{li2007eos}, \cite{zhang2007expert}.  However,
automated expert discovery is still challenging for a variety of reasons.
First, effectively collecting or generating a meaningful expertise list for
each individual is not a straightforward task.  Second, given a query term $q$,
it is not obvious how to rank the potential experts who have the need expertise skills $q$.
Third, combining experts based on terms $q'$ that are
synonyms of or similar to $q$ is not straightforward.

We propose ExpertSeer\footnote{\url{http://expertseer.ist.psu.edu/}},
an open source keyphrase-based recommender system for expert and related topic
discovery.  ExpertSeer approaches the three challenges described above in a principled
way.  Based on a given digital library and accessory resources, such as
Wikipedia, ExpertSeer generates keyphrases from the title and the abstract of
each document in the digital library.  These keyphrases are further utilized to
infer the authors' expertise and to relate similar terms.  To rank the experts
in a given field, the system relies on Bayes' rule to integrate the
relevance and authors' authority on a given field.

In order to demonstrate the generality of our framework, we have used
ExpertSeer to build two expert recommender systems: one for computer scientists
(CSSeer\footnote{\url{http://csseer.ist.psu.edu/}}) and another for chemists
(ChemSeer\footnote{\url{http://chemseer.ist.psu.edu/}}).  The initial
experimental results for CSSeer are promising.  Our system was able to assign
high quality keyphrases to more than $95\%$ of the documents.  In addition, the
experts recommended are mostly prestigious scholars in the relevant domain.
Based on a widely used expert list, our system outperforms two state-of-the-art
expert recommenders, ArnetMiner\footnote{\url{http://arnetminer.org/}} and
Microsoft Academic Search (MAS)\footnote{\url{http://academic.research.microsoft.com/}},
in terms of Precision-at-$k$ ($k=3,5,10$).  Furthermore, users may take advantage
of the related keyphrase list to compile a more comprehensive list of experts,
since state-of-the-art expert recommenders still generate divergent recommendations,
as demonstrated in our experiments.

This work makes the following contributions.
\begin{enumerate}
  \item We designed ExpertSeer, an open source general framework for expert
    recommendation and related keyphrase discovery based on a digital library.
    Institutes may utilize the system to build an expert recommender based on their
	own internal or personal collection of documents.  To our knowledge, ExpertSeer
    is the first open source framework for expert recommendation for a scholarly
    digital library.
  \item We applied the generic framework to two different disciplines, namely
    computer science and chemistry.  The system is highly scalable and efficient
    in managing digital libraries with millions of documents and authors.
  \item Using CSSeer, we compare empirically the performance
    of state-of-the-art expert recommenders.  The results show that current
    expert recommenders still have a moderately divergent suggested list.  Based
    on a publicly available dataset, our system outperforms the others in terms
    of Precision-at-$k$ ($k=3,5,10$).
  \item We have validated that Wikipedia can be a promising keyphrase candidate
    source on keyphrase extraction of academic documents for large size digital
    libraries.
\end{enumerate}

The rest of the paper is organized as follows.  In Section~\ref{sec:rel}, we
review previous works on keyphrase extraction, related term discovery, and
expert recommendation.  Section~\ref{sec:method} introduces our methods for
keyphrase extraction, related keyphrase compilation, expert recommendation, and
expertise list compilation.  Section~\ref{sec:exp} shows the experiments,
evaluation metrics, and results.  Several case studies are presented in
Section~\ref{sec:case}.  Finally, a summary and description of future work
appear in Section~\ref{sec:conc}.

\section{Related Work} \label{sec:rel}

ExpertSeer automatically extracts keyphrases from documents.  Based on these
keyphrases, ExpertSeer discovers related phrases and builds the expert list.
In the section, we review previous works on keyphrase (or keyword) extraction,
related phrase compilation, and expert recommendation techniques.

Automatic extraction of keyphrases in documents has become quite popular.
Traditional automatic keyphrase extraction usually consists of two stages:
candidate keyphrase selection and keyphrase identification from these
candidates~\cite{jones2002automatic}, \cite{nguyen2007keyphrase}, \cite{witten1999kea}.  The
candidate selection process would include many potential keyphrases to achieve
a higher recall, but by randomly increasing the size of candidate keyphrases comes
the risk of a lower precision and hurting analysis efficiency.  One
popular method to identify candidate keyphrases is exploiting part-of-speech (POS)
taggers to extract nouns or noun phrases as candidates~\cite{kim2009re},
\cite{nguyen2007keyphrase}.  Another possible alternative is to include frequent
$n$-grams in the candidate list~\cite{treeratpituk2010seerlab}.  However, all
of these methods tend to include many trivial and relatively vague terms, such
as ``study'', ``method'', ``model'', etc.  As a result, the performance relies
heavily on the keyphrase identification process, which is usually a supervised
learning process that typically relies heavily on lexical and syntactic
features, such as term frequency, document frequency, and term
locations~\cite{nguyen2007keyphrase}, \cite{treeratpituk2010seerlab}.  Recently,
methods that utilize features from the Wikipedia
corpus~\cite{grineva2009extracting}, \cite{mihalcea2007wikify} were shown to select
better keyphrases compared to pure TF-IDF based
methods~\cite{grineva2009extracting}.  However, these learning methods require
a large number of training samples to learn a representative model.  In contrast
to these approaches, ExpertSeer uses only simple stemming and matching
as opposed to learning from Wikipedia pages and yet still efficiently extracts high
quality terms with high recall from scientific literature and is much more efficient.

To discover semantically related terms, the most popular way is to use
well-known lexical databases, such as WordNet~\cite{pedersen2004wordnet} and
FrameNet~\cite{ruppenhofer2006framenet}.  However, these databases usually have
poor coverage for terms in science and engineering fields~\cite{turney2001mining}.
The co-appearance of words or mentions have been shown to be a good indicator of
topic relevance in practice~\cite{zhou2009assessment}.  Recently, researchers have
resorted to Wikipedia for related term extraction.
WikiRelate!~\cite{strube2006wikirelate} employed text distance and category
path distance between pages to define the relevance between words in the page
title.  However, WikiRelate! was limited to comparing unigrams.  Gabrilovich and
Markovitch~\cite{gabrilovich2007computing} transformed terms into a higher
dimensional space of concepts derived from Wikipedia.  The hyperlink structure
of Wikipedia was shown to be an effective measure of relatedness between
terms~\cite{witten2008effective}.  Milne and Witten showed the practicality of
identifying key concepts from plain text using
Wikipedia~\cite{milne2008learning}.  Following this line, we are the first to
combine Wikipedia pages with scientific literature to infer the relatedness
between scientific terms based on Bayes' rule.

Today, expert discovery continues to be a problem of interest.  The problem
involves several practical issues, including author name disambiguation,
profiling user data (such as contact information and expertise list), defining
an expert ranking function, etc.  Microsoft's Libra project (now renamed
Microsoft Academic Search) performed name disambiguation and user profiling by
identifying and extracting information from every researcher's
homepage~\cite{zhu2007webpage}.  A similar approach was also applied by
ArnetMiner~\cite{tang2008arnetminer}.  To identify experts, authors were
associated with the text or topics of their publications based on various
models~\cite{balog2006formal}, \cite{fang2007probabilistic}, \cite{yukawa2001expert},
\cite{li2011approach}.  Numerous
studies suggested associating authors with not only published papers but also
conferences or journals~\cite{tang2008arnetminer}.  To infer the quality of
academic articles or the authority of authors, citation-based indices were shown to be good
indicators~\cite{deng2008formal} \cite{hu2011analyzing}.  In addition to publications, email
communication was also utilized when suggesting experts within an
enterprise~\cite{campbell2003expertise}.  Several studies performed expert
finding by utilizing social network link structure, including propagation based
approaches~\cite{deng2011probabilistic}, \cite{li2007eos},
\cite{zhang2007expert}, \cite{fan2013expfinder}, a constraint regularization
based approach~\cite{deng2012modeling}, and a PageRank-like
approach~\cite{gollapalli2011ranking}.  Expert finding has also been applied to
social media, such as forums or community question answering portals, to
recognize reliable users and contents~\cite{bian2009learning}, \cite{pal2012exploring}.

Initial experimental results of CSSeer~\cite{chen2013csseer} gave a comparison with ArnetMiner and MAS.
In this paper, we report several improvements and new functionalities of the system, including the
ranking function, the related keyphrase extraction, the expertise generation, and
practical computational issues.  We also conducted several experiments and case
studies.

\section{Methodology} \label{sec:method}

We now introduce the methodology of keyphrase extraction, expert recommendation and
ranking function, expertise list compilation, and related phrase discovery of
the ExpertSeer framework.  We also discuss the scalability and the incremental
updating procedures, which are important but often overlooked issues for
growing or changing digital libraries.

\subsection{Keyphrase Extraction}\label{sec:keyphrase_extract}

Similar to most state-of-the-art keyphrase extractors, ExpertSeer applies a
two-stage approach, namely candidate keyphrase selection and keyphrase
identification from candidates.  To effectively collect meaningful academic
terms as candidates, we resort to two sources: Wikipedia pages and the
documents in a given digital library.

ExpertSeer employs Wikipedia to effectively collect meaningful academic terms
as candidates.  Since category information and pages within a category on
Wikipedia are compiled manually, they have highly reliable semantic meaning.
The categorization of Wikipedia is utilized to collect terms related to our
target domain.  Take CSSeer for example, the crawler started from the category
``computer science'' and retrieved all pages in the category (depth 0), all the
pages in the sub-category of computer science (depth 1), up to all pages in the
depth 3 category.  By a similar manner, all pages under the category ``statistics''
and the category ``mathematics'' up to depth 2 were extracted, since computer
scientists use many statistical and mathematical techniques.  The titles and
the hyperlink texts from the introduction paragraphs of these pages
were retrieved as possible keyphrase candidates.  Since the titles and hyperlink
texts are edited by users, they are usually rich with meaningful semantics.  Trivial
or vague terms, such as ``study'', ``method'', ``model'', which were usually
selected as keyphrase candidates by previous methods, are unlikely to be
selected.  To increase the recall, the bigrams, trigrams, and quadgrams that
appear at least 3 times in the titles of the documents in the digital library
are also included in the candidate list.  Compared
to~\cite{treeratpituk2010seerlab}, which selects frequent $n$-grams as
keyphrase candidates, a Wikipedia based method is better because semantically
meaningful terms can be naturally included in the candidate list.  In addition, it
is not straightforward to specify the maximum value of $n$ for $n$-gram based
methods.  A small value excludes longer terms (e.g., ``strong law of large
numbers''), but a large value makes the matching process time consuming and may
inevitably include several questionable terms.

To identify keyphrases for each document in the digital library, our framework
constructs a trie from all the collected keyphrase candidates, and compares all
the titles and abstracts with the keyphrase candidates based on the trie
structure, which can efficiently perform the longest-prefix-matching
lookup~\cite{treeratpituk2010seerlab}.  If a match is found, the matched term is
selected as one keyphrase of the document.  As shown later in
Section~\ref{sec:wiki-coverage}, such a method works effectively for more than
$95\%$ of the documents in our tested corpus.  Compared to supervised learning
based keyphrase identification approaches~\cite{nguyen2007keyphrase},
\cite{treeratpituk2010seerlab}, the stemming and matching method is not dependent on
the training data.  In addition, it is very simple and efficient in practice.

\subsection{Expert Ranking}\label{sec:exp_rank}

ExpertSeer discovers experts of a given term based on Bayes' rule.  The
model naturally integrates textual relevance and quality of the authors'
published papers within a unified framework.

Below, we start by introducing the case where a query term appears in the
keyphrase candidate (which is compiled in advance, as introduced in
Section~\ref{sec:keyphrase_extract}).  In this scenario, our system computes the
required information offline so that it can efficiently respond to users'
queries.  Next, we will show how to approximate the result when the query term
does not appear in the candidate list.

\subsubsection{The Query Term Appears in the Candidate List}
\label{sec:query-in-candidate}

Similar to~\cite{balog2006formal} and~\cite{deng2008formal}, we define the
problem by a probability model: what is the conditional probability $p(a|q)$
that an author $a$ is an expert given a query $q$?  By Bayes' rule, $p(a|q)$
can be written as follows.

\begin{equation}\label{eq:expert_prob_bayesian}
p(a|q) = \frac{p(a, q)}{p(q)} \propto p(a,q),
\end{equation}
where the denominator term $p(q)$ can be disregarded because $q$ is fixed by
the time $p(a|q)$ needs to be determined.

To introduce the set of documents $D$ to the model,
Equation~\ref{eq:expert_prob_bayesian} is rewritten into the following form.

\begin{equation} \label{eq:expert_prob_with_doc}
\begin{array}{rcl}
p(a|q) \propto p(a,q) & = & \sum_{\forall d \in D} p(d) p(a, q | d) \\
 & = & \sum_{\forall d \in D} p(d) p(q | d) p (a | q, d) \\
 & = & \sum_{\forall d \in D} p(d) p(q | d) p (a | d),
\end{array}
\end{equation}
where the last equality holds since an author $a$ is conditionally independent to
a query $q$ given the document $d$.

We interpret each term in Equation~\ref{eq:expert_prob_with_doc} below.

The term $p(d)$ represents the probability that document $d$ is an important
document.  This can be inferred by several possible metrics, such as the
citation counts, the number of downloads, the reputation of the published
conference or journal, graph-based algorithms (e.g., PageRank-like algorithms),
or a combination of these factors.  Earlier studies show that the number of
citations is positively related to the number of
downloads~\cite{watson2009comparing} and graph-based
measures~\cite{ding2009pagerank}.  Thus, we simply utilize the citation counts to
infer the quality of a paper.
In addition, Deng et al.\ showed that the logarithm of the number of
citations is a better indicator of paper quality among several
alternatives~\cite{deng2008formal}.  Thus, ExpertSeer sets the value of $p(d)$
as the logarithm of the number of citations of $d$.

The term $p(q|d)$ is the probability that $q$ is relevant given $d$.  We set
$p(q|d)$ as a variation of the language model, as shown by
Equation~\ref{eq:bag-of-phrases-model}.  However, other textual relevance
measures, such as TF-IDF and BM25, can be applied too.

\begin{equation} \label{eq:bag-of-phrases-model}
p(q|d) = \frac{|d|}{|d|+\mu} \cdot \frac{c(q, d)}{|d|} + \left(1 - \frac{|d|}{|d|+\mu}\right) \cdot \frac{c(q, D)}{|D|},
\end{equation}
where $|d|$ is the total counts of phrases, not words, in the document $d$,
$\mu$ is the Dirichlet smoothing factor, which is used to prevent
under-estimating the probability of any unseen phrases in
$d$~\cite{zhai2001study}, $c(q, d)$ is the frequency of $q$ in $d$, $|D|$ is the
number of phrases, not words, in the corpus $D$, and $c(q, D)$ is the frequency
of $q$ in $D$.

Equation~\ref{eq:bag-of-phrases-model} is different from the classic language
model for the following reasons.  The traditional language model represents
documents based on the bag-of-words (BOW) assumption, which treats each word as
a basic token and assumes independence between words.  Our method represents a
document by a bag-of-\emph{phrases} model, which may capture the contexts of a
document beyond the granularity of a word.  Our system identifies most of the
phrases in the documents based on the earlier compiled keyphrase candidate list.
If certain texts do not match any phrases in the list, our system tokenizes this
piece of texts into words, and applies the classic language model.  Similarly,
when the query term $q$ is an $n$-gram formed by words $w_1 w_2  ...  w_n$ ($n >
1$), the classic language model has the independence assumption so that $p(q|d)
= p(w_1, w_2, \ldots, w_n | d) = p(w_q|d) p(w_2|d) \ldots p(w_n|d)$.  In
practice, however, $w_1, \ldots, w_n$ depends on others and the sequence of
$w_1,\ldots, w_n$ matters.  For example, when we read ``support vector'', it is
very likely that the next word is ``machine'', since ``support vector machine''
as a whole is a complete phrase.

The term $p(a|d)$ accounts for the contribution of an author $a$ given a
document $d$.  One possible choice is to divide the contribution equally by the
number of authors, as applied in~\cite{deng2008formal}.  Thus, $p(a|d) = 1/n_d$
if $a$ is one of the $n_d$ number of authors of $d$ and 0 otherwise.  One could
also suggest other models such as giving more credits to the first author than
the other authors.  For simplicity, we use an indicator function to define the
value: $p(a | d) = 1$ if $a$ is an author of $d$ and $p(a | d) = 0$ otherwise.

\subsubsection{The Query Term does not appear in the Candidate
List}\label{sec:query_not_in_candidate}

For a term $q$ in the candidate keyphrase list, the expert score $p(a|q)$ can be
computed offline, as shown in Equation~\ref{eq:expert_prob_with_doc}.  However,
users may submit a query term $q'$ that is not included in the candidate list.
Calculating $p(a|q')$ in real time is impractical, since we need to accumulate
$p(d)p(q'|d)$ for all $d$ in each author's publications.

One na\"{i}ve way to bypass the problem is to aggregate all of the documents for
each author and build an inverted index to map words to authors.  However, the
inverted index has no document information.  As a result, the quality of the
documents is not included in the model.  In such a setting, the recommender could potentially return authors who wrote several mediocre documents on topic $q'$.

To solve this problem, we reformulate Equation~\ref{eq:expert_prob_with_doc} as
the following (assuming $q'$ is not in the candidate list).

\begin{equation} \label{eq:expert_prob_partial}
\begin{array}{rcl}
p(a|q') & \propto & \sum_{\forall d \in D} p(d) p(q' | d) p (a | d) \\
       & = & \sum_{\forall d \in D_1} p(d) p(q' | d) p(a|d) \\
       &   & + \sum_{\forall d \in D_2} p(d) p(q' | d) p(a|d) \\
       & \approx & \sum_{\forall d \in D_1} p(d) p(q' | d) p(a|d),
\end{array}
\end{equation}
where $D_1 \cup D_2 = D$, $D_1 \cap D_2 = \emptyset$, and $D_1$ is composed of
$n$ documents with the highest $p(q', d)$ values in $D$.  Thus, only the authors
of the documents with top-$n$ $p(q', d)$ scores are integrated and ranked.  The
documents with lower $p(q', d)$ scores contribute less to the score of $p(q',d)$
and are left out.  The values of $p(d)$ and $p(a|d)$ are calculated as
introduced in Section~\ref{sec:query-in-candidate}.  Since the query term $q'$
is not in the keyphrase candidate list, we cannot apply
Equation~\ref{eq:bag-of-phrases-model} to obtain $p(q'|d)$ directly.  Instead,
we use Equation~\ref{eq:modified-bag-of-words} to get $p(q'|d)$.

\begin{equation} \label{eq:modified-bag-of-words}
p(q'|d) = \prod_{\forall w \in q'} \left(\frac{|d|}{|d|+\mu} \cdot
    \frac{c(w,d)}{|d|} + \left(1-\frac{|d|}{|d|+\mu}\right) \cdot
    \frac{c(w,D)}{|D|}\right),
\end{equation}
where $w$'s are the words in $q'$.  The equation is different from the classic
language model in that $|d|$ and $|D|$ are the total number of phrases, not
words, in $d$ and $D$, respectively.

To efficiently discover the $n$ documents with top $p(q',d)$
values, the Apache Solr\footnote{\url{http://lucene.apache.org/solr/}} system is
employed to build full text index and perform function queries.

\subsection{Expertise List Compilation and Ranking}

When a user queries an author $a$, the system shows the expertise list of $a$.
This section introduces the compilation as well as the ranking function of the
expertise list.

Similar to the expert ranking method, we formally define the problem by a
conditional probability distribution: what is the conditional probability
$p(t|a)$ that a term $t$ is one research expertise given the author $a$?
Similar to Equation~\ref{eq:expert_prob_with_doc}, it can be derived as
follows.

\begin{equation} \label{eq:term_prob_of_author}
p(t|a) \propto p(t, a) = \sum_{\forall d \in D} p(d) p(t | d) p (a | d).
\end{equation}

The terms $p(d)$, $p(t|d)$, and $p(a|d)$ are calculated by the same method
introduced in Section~\ref{sec:query-in-candidate}.

\subsection{Related Phrase Compilation}

Different authors may use different terms to describe the same or similar ideas.
For example, ``logistic regression'' is also known as ``logit model''.  When
searching for experts of ``logistic regression'', authors who usually use
``logit model'' to refer to ``logistic regression'' may not be considered as
experts by an expert recommender.  In addition, we may want the system to return
experts of relevant areas as well.  For example, when searching for experts of
``logistic regression'', we may also be interested in knowing the experts of
``binary classifier'' and ``multi-nominal logistic regression''.

To include the experts of relevant topics, ExpertSeer provides a list of
related keyphrases of the query term.  Thus, users may browse through the
experts of the relevant topics to compile a more comprehensive expert list.  To
ensure that the list includes only non-trivial terms, the list is a subset of the
keyphrase candidates.

A na\"{i}ve way to infer the relatedness between two terms is the co-appearance
frequency.  However, such a method favors the high frequency terms, i.e., the
higher frequency terms tend to be related to every other term.

Instead of counting co-appearance frequency, CSSeer exploits Bayes' rule to
discover related phrases.  More formally, given a query term $t$, the
relatedness score of another term $s$ to $t$ is given by $p(s|t)$: the
conditional probability that $s$ is relevant to a document given that $t$ is
relevant to the document.  The value of $p(s|t)$ is derived by the following
equation.

\begin{equation} \label{eq:term_rel_score}
\begin{array}{rcl}
p(s|t) & \propto & p(s, t) \\
       & = & \sum_{\forall d \in D} p(d) p(s,t | d) \\
       & = & \sum_{\forall d \in D} p(d) p(t|d) p(s | t,d) \\
       & = & \sum_{\forall d \in D} p(d) p(t|d) p(s|d)
\end{array}
\end{equation}

The terms $p(t|d)$ and $p(s|d)$ are calculated by
Equation~\ref{eq:bag-of-phrases-model}.  The term $p(d)$ is the probability that
$d$ is an important document.  A document $d$ is usually more carefully edited
if it is more authoritative, and thus the wording is usually more precise.
Moreover, other authors are more likely to follow the wording behavior used in
$d$.  As a result, we should assign a higher relevance score to two terms
appearing in a more authoritative document.  The value of $p(d)$ can be inferred
based on several factors, such as citation counts and download counts, as
suggested in Section~\ref{sec:exp_rank}.

\subsection{Incremental Updating and Scalability}

To support a live digital library that includes new documents over time,
incremental updating is very important.  For ExpertSeer to import new documents
and perform incremental updating, the metadata, citation list, and the
keyphrases are extracted when a new document is imported.  ExpertSeer updates
the following records according to the extracted information.  First, the system
may add an author to the author list if identified as a new author.  Second, the
system utilizes the extracted keyphrases to update the authors, expertise list,
and the related keyphrase information.  Finally, the citation counts of the
cited papers are increased.  ExpertSeer accomplishes these updates easily,
given that it indexes the authors, expert list, keyphrase relationship, and paper
information.

ExpertSeer is highly scalable.  CSSeer, one of the expert recommender built
from ExpertSeer, currently handles over $1,000,000$ documents and over
$300,000$ distinct authors efficiently.

\section{Experiments} \label{sec:exp}

We conducted extensive experiments on the system from several different aspects.
We compared the lists of the top-$n$ returned experts from CSSeer, ArnetMiner,
Microsoft Academic Search (MAS), and GS*, a system we used to simulate Google
Scholar's ranking function.  We build GS* to simulate Google Scholar's ranking
function on the top of CiteSeerX's dataset, because Google Scholar does not
provide APIs for users to efficiently query a long list of queries.  We also
investigated the performance of the Wikipedia based keyphrase extractor.

\subsection{Consensus among Different Expert Recommenders}\label{sec:consensus}

\begin{table}[tb]
\begin{center}
\caption{\label{tab:dm_cmp_example}The top 10 experts of ``data mining'' returned by CSSeer, ArnetMiner, and Microsoft Academic Search (MAS).  Scholars appearing in the top 3 by at least two of them are highlighted by $\dag$; scholars appearing in the top 5 by at least two of them are highlighted by $\ddag$; scholars appearing in the top 10 by at least two of them are highlighted by $*$.  $S@n$: consensus score for the top $n$ returns.}
\begin{tabular}[t]{c|ccc}
\hline
Rank & CSSeer & ArnetMiner & MAS \\
\hline \hline
1 & Jiawei Han $\dag \ddag *$ & Jiawei Han $\dag \ddag *$ & Jiawei Han $\dag \ddag *$ \\
2 & Salvatore J. Stolfo & Philip S. Yu $\dag \ddag *$ & Philip S. Yu $\dag \ddag *$ \\
3 & Mohammed J. Zaki $\dag \ddag *$ & Mohammed J. Zaki $\dag \ddag *$ & Tzung-Pei Hong \\
4 & Osmar R. Zaiane & Christos Faloutsos $*$ & Yong Shi \\
5 & Maciej Zakrzewicz & Jian Pei & Shusaku Tsumoto \\
6 & Krzysztof Koperski & Heikki Mannila & Alex Alves Freitas \\
7 & Marek Wojciechowski & Rakesh Agrawal & Andrew Kusiak \\
8 & Christos Faloutsos $*$ & Charu C. Aggarwal & Mohammed Javeed Zaki \\
9 & Wei Wang & Raymond Ng & Vipin Kumar \\
10 & Srinivasan Parthasarathy & Usama M. Fayyad & Xin-Dong Wu \\
\hline \hline
$S@3$ & 2 & 3 & 2 \\
\hline
$S@5$ & 2 & 3 & 2 \\
\hline
$S@10$ & 3 & 4 & 2 \\
\hline
\end{tabular}
\end{center}
\end{table}

\fig{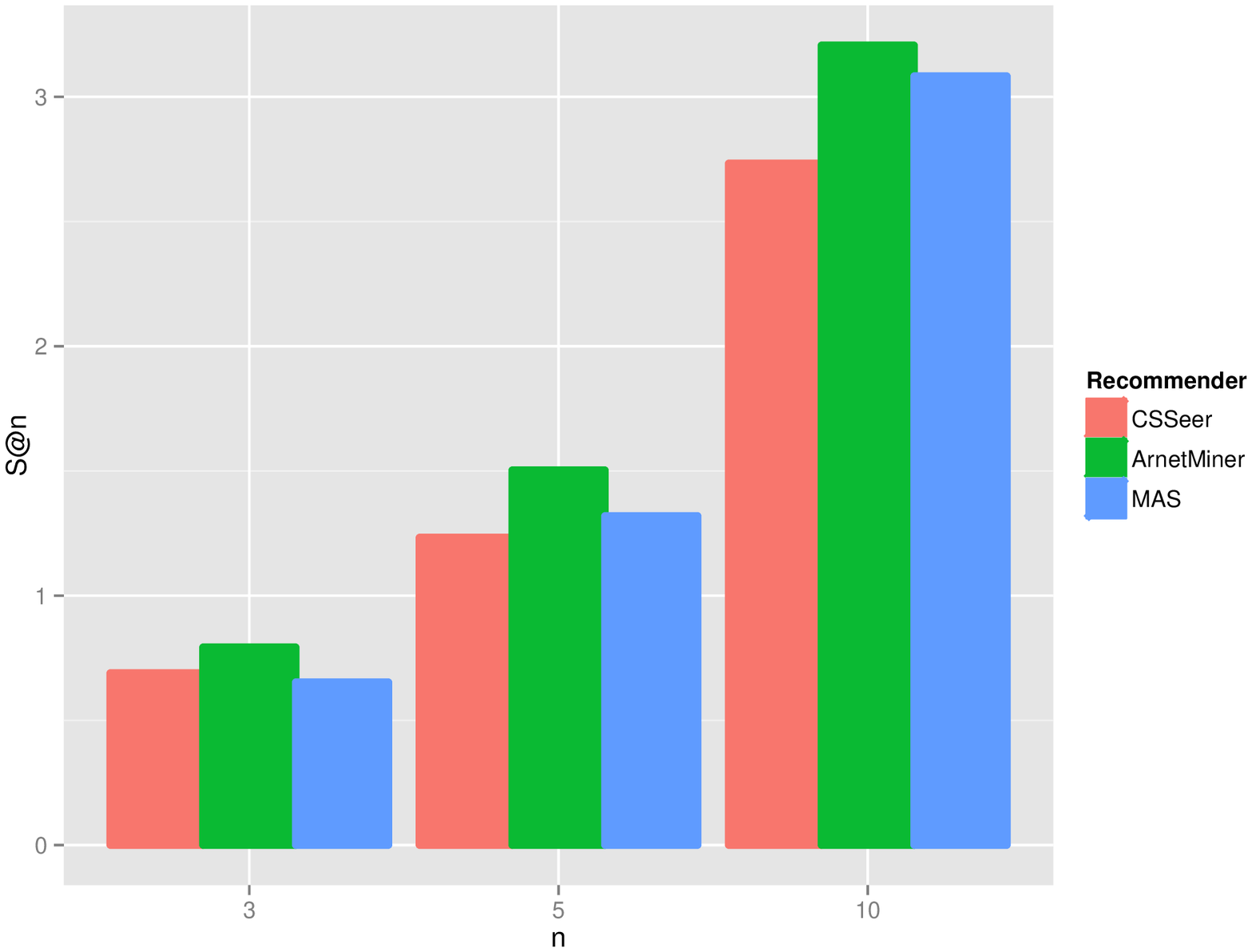}
    {Consensus scores $S@n$ ($n=3,5,10$) using 1,000 benchmark queries.}
    {fig:consensus_score}{2.5}

Evaluating a recommender system usually requires an extensive user study.  To
evaluate an expert recommender system, it is even more difficult since the
evaluators need to have sufficient domain knowledge in order to identify the experts of a given topic.  Although CSSeer focuses mainly on Computer Science, the
sub-domains are still very diverse, ranging from software engineering, data
management, applications, to compiler, architecture, and system chip design.
As a result, it is very difficult to rely on a small number of individuals to
evaluate the expert list in several different domains.

To evaluate the performance of CSSeer at a large scale, we compared the expert
list returned by CSSeer with two other expert recommender systems, namely
ArnetMiner and Microsoft Academic Search, in terms of their recommending
consensus.  Specifically, we compared the overlap of the top $n$ returned
experts of the three systems ($n=3,5,10$).  We measured only the overlap of the
returns instead of using position based measurements, such as discounted cumulative
gain~\cite{jarvelin2002cumulated} and expected reciprocal
rank~\cite{chapelle2009expected}.  The reason for this is that given a query
term, the top returned names by all three systems are mostly prestigious
researchers.  Asking an evaluator to differentiate who might be more
knowledgeable among a list of reputable researchers is not an easy task and is
very likely to be a biased evaluation.

To quantify the measurement, we define the consensus score $S@n$ of one expert
recommender system $e_i$ to the other systems $e_1, \ldots, e_{i-1}, e_{i+1},
\ldots e_m$ in Equation~\ref{eq:consensus_at_n}~\cite{chen2013csseer}.

\begin{equation} \label{eq:consensus_at_n}
S@n \equiv \left|\bigcup_{\forall k \neq i}\left(r_i^{(n)} \cap r_k^{(n)}\right)\right|,
\end{equation}
where $r_i^{(n)}$ is the set of the top $n$ returns of the $i$th recommender
$e_i$, and the $| \cdot |$ function returns the set length.

To make the concept of consensus score clearer, we show $S@n (n=3,5,10)$ for
the three systems using a query term ``data mining''.  The top 10 names
returned by these systems are shown in Table~\ref{tab:dm_cmp_example}.  Among
the returned names of CSSeer, 3 of them (Jiawei Han, Mohammed J. Zaki, and
Christos Faloutsos) appear in at least one of the other two system's top 10
list.  Thus, $S@10$ for CSSeer would be 3.  In a similar manner, we can
calculate $S@10$ for ArnetMiner and MAS as 4 and 2 respectively.  Note that
although Christos Faloutsos ranked 4th by ArnetMiner, he cannot be counted when
calculating $S@5$ for ArnetMiner, because the name neither appears in the top 5
returned names of CSSeer nor MAS.

The computation of consensus scores involves no user evaluation, and would thus
be amenable to automation of the evaluation process to a large number of
queries.  However, there is a problem in practice: different expert recommender
systems may record the same expert with different name variations~\cite{chen2013csseer}.  For
example, Dr.\ Michael I.  Jordan at the University of California Berkeley is
recorded as ``Michael I.  Jordan'' in both CSSeer and MAS but is ``M. I.
Jordan'' in ArnetMiner.  Dr.\ ChengXiang Zhai at University of Illinois at
Urbana-Champaign is stored as ``ChengXiang Zhai'' in both CSSeer and ArnetMiner
but is ``Cheng-xiang Zhai'' in MAS.  Therefore, na\"{i}vely regarding names as
strings and performing string matching could generate misleading results.  To
automate the name disambiguation, we normalized each returned name by lower-casing
each letter and keeping only the last name and the first letter of the first
name.  Thus, ``Michael I. Jordan'' and ``ChengXiang Zhai' are normalized as ``m
jordan'' and ``c zhai'' respectively.  Since only the top $n$ returned names
are compared, it is less likely that two experts of the same field share the
same last name and similar first names.

We compared $S@n$ ($n=3,5,10$) of the three systems for 1,000 benchmark queries.
Although we could use the relevant judgments provided by ArnetMiner
directly\footnote{\url{http://arnetminer.org/lab-datasets/expertfinding/}}~\cite{yang2009expert2bole},
the number of terms is very small and these terms are mainly of the artificial
intelligence, data mining, and information retrieval domains.  In the hope of
covering diverse sub-domains of Computer Science, we intentionally included
terms of diverse topics, including hardware (such as ``VLSI''), low level
machine concepts (such as ``compiler'' and ``virtual machine''), software
development (such as ``programming language'', ``data structure'', and
``software engineering''), statistical techniques (such as ``nonparametric
statistics'' and ``markov chain monte carlo''), data mining techniques (such as
``conditional random fields'' and ``support vector machine''), and so on.  Thus,
the 1,000 benchmark queries of terms are diverse and contain both broad and
narrow topics.

The consensus scores of the benchmark queries on the three systems are shown in
Figure~\ref{fig:consensus_score}.  As one can see, the average consensus scores
$S@n$ ($n=3, 5, 10$) are low for all three expert recommenders.  Specifically,
on average only $0.653$ to $0.793$ names out of the top 3 returned of one
system are overlapped with at least one of the other two systems.  For the top
5 returns, the numbers of overlapping names are also small, on average ranging
from $1.233$ to $1.503$.  For $n=10$, the number of overlapped names are
ranging from $2.733$ to $3.207$.  This suggests that the current
state-of-the-art expert recommender systems still have divergent opinions.
Relying on only one expert recommender system may obtain a biased expert list.

\subsection{Precision Comparison of Different Expert Recommenders}

\begin{table}[tb]
\begin{center}
\caption{\label{tab:prec_at_k_cmp}Precision at $k$ ($P@k, k=3, 5, 10$) for
different expert recommenders, based on the expert list
given in~\cite{tang2008arnetminer}}
\begin{tabular}[t]{c|ccc}
\hline
 & $P@3$ & $P@5$ & $P@10$ \\
\hline \hline
CSSeer & \textbf{0.6667} & \textbf{0.7077} & \textbf{0.5538} \\
ArnetMiner & 0.6410 & 0.6308 & \textbf{0.5538} \\
MAS & 0.6154 & 0.6 & 0.5308 \\
GS* & 0.1538 & 0.2308 & 0.2462 \\
\hline
\end{tabular}
\end{center}
\end{table}

The consensus comparison of the systems discussed in last section shows that in
many cases different systems give preference to different experts.  However, it
is difficult to compare the quality of different expert recommenders because
there is no base standard for reference.  To further investigate their
performance, user evaluation is inevitably needed.

Instead of conducting expensive user study, we obtain the relevant judgments
provided in~\cite{tang2008arnetminer} as the golden standard for expert list
comparison.  We selected Precision-at-$k$ ($P@k$) as the evaluating metric.
Although position aware metrics, such as Discounted Cumulative Gain and Mean
Reciprocal Rank, can be applied, they are not selected because such measures
are very likely to be biased for expert list evaluation, as discussed
in Section~\ref{sec:consensus}.

We compared the returned names of the three systems (CSSeer, ArnetMiner, and
Microsoft Academic Search) and GS*, a system we built to simulate Google
Scholar's ranking function.  Google Scholar asks authors to manually input up to
5 phrases to represent their research expertise.  When a user submits a query
term $q$, Google Scholar retrieves all authors who lists $q$ as their expertise,
and rank these authors by the total number of citations they have received.  GS*
simulates Google Scholar's behavior as follows: it first retrieves all authors
who published papers related to the query term (based on the documents collected
by CiteSeerX), and then ranks these authors by their total citation counts.  This
approach considers both authors' research interest and authority.  However, the
ranking function is only based on the total number of citations.  As a result,
if an author published many high quality papers in area 1 but only several
mediocre papers in area 2, the author would still ranked very high when the
query term is related to area 2.

Table~\ref{tab:prec_at_k_cmp} shows the evaluation results of these systems.
When retrieving experts by relevancy and ranking the result by authority, as GS*
does, the performance is mediocre.  All three state-of-the-art systems (CSSeer,
ArnetMiner, and Microsoft Academic Search) perform reasonably well for the
top-3, top-5, and top-10 returns, because the ranking function includes not only
the relevance between the query term and the authors' research fields but also
the authority of the author in regards to this term.  Among the three expert
recommenders, our proposed system, CSSeer, on average performs best.  The
average scores of $P@3$, $P@5$, and $P@10$ on the benchmark queries are
$0.6667$, $0.7077$, and $0.5538$ respectively.

We expect CSSeer to perform better than the other two for the following two
reasons.  First, both ArnetMiner~\cite{tang2011topic} and
MAS\footnote{\label{foot:mas-faq-url}\url{http://academic.research.microsoft.com/About/help.htm}}
seem to treat each word as an independent token.  However, a term (e.g.,
``support vector machine'') may consist of a set of words.  CSSeer is very
likely to group and index the entire term as one token, since such a term is
highly likely to be included in the keyphrase candidate list compiled from
Wikipedia and the frequent $n$-grams in the titles of the papers in the given
corpus.  Second, CSSeer probably assigns a more appropriate authority score to
authors of a given query.  For a set of authors who have published papers
related to a query $q$, ArnetMiner employees a propagation-based approach on the
coauthorship network to rank these authors~\cite{li2007eos}, \cite{zhang2007expert}.
Specifically, ArnetMiner first claims authors who wrote several papers related
to $q$ as potential experts, and then assumes that authors who have coauthored
with potential experts are more likely to be experts as well.  Such a method,
however, does not incorporate the citation information, which is usually a good
indicator of the quality of a paper.  MAS computes Field Rating -- the rating of
authors on a field -- of each author on some terms in
advance\footnote{See footnote~\ref{foot:mas-faq-url}}.  However, when a query term is not
in the pre-computing list, the ranking function seems to be similar to Google
Scholar.  As a result, an author who is highly authoritative in one area may
dominate the results of another area in which she is less authoritative.

\subsection{Coverage of Wikipedia Based Keyphrase Candidates}
\label{sec:wiki-coverage}

%

\figtwo{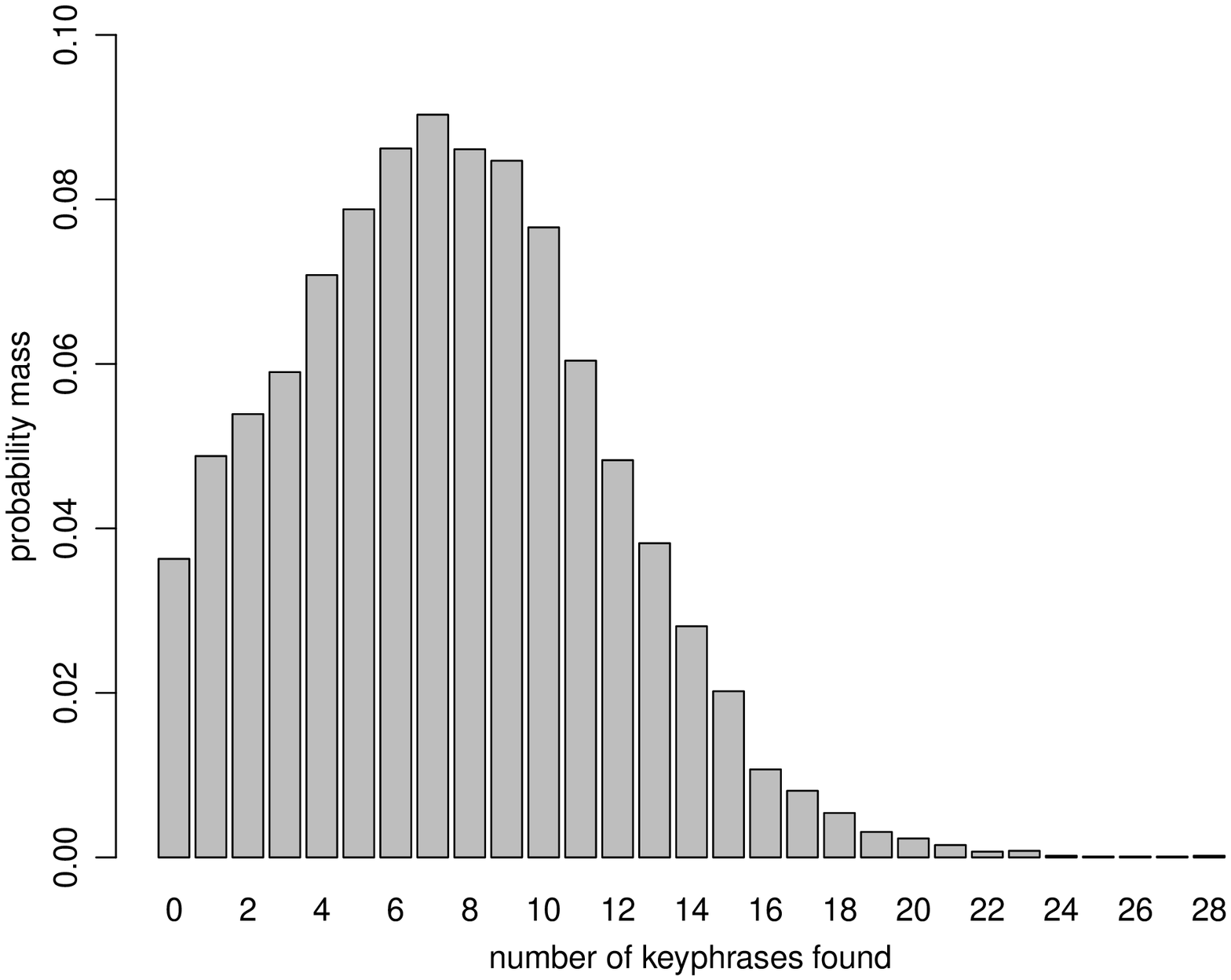}{Set A: $10,000$ randomly selected documents}{fig:keyphrase_per_doc_no_prune}
       {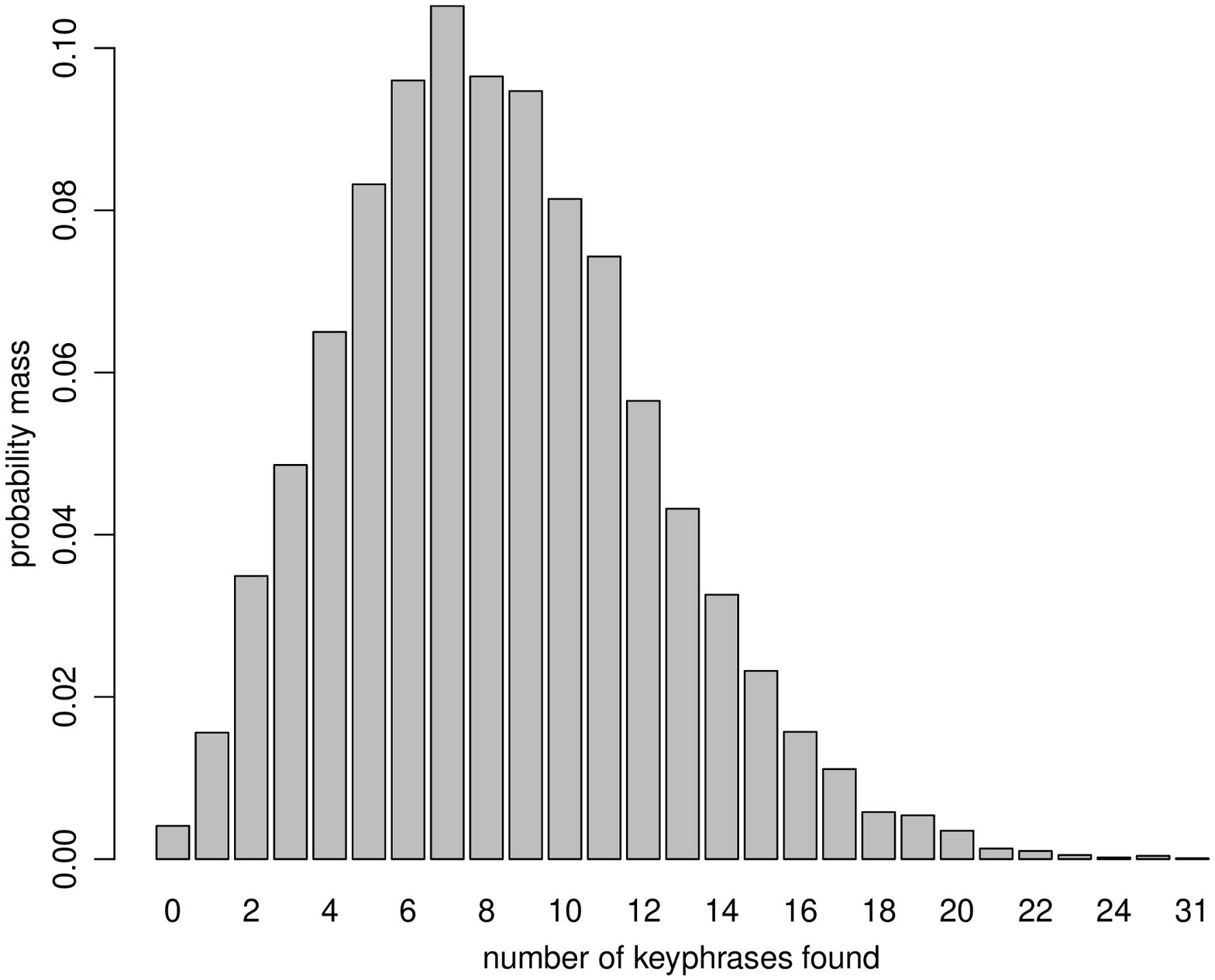}{Set B: $10,000$ randomly selected documents whose titles contain least 4 words and abstracts contain at least 20 words}{fig:keyphrase_per_doc_min_length}
       {Empirical probability mass function of number of keyphrases found in title and abstract for a document in CiteSeerX.}{fig:keyphrase_per_doc}

\begin{table}[tb]
\begin{center}
\caption{\label{tab:keyphrase_per_doc}Statistics of the number of keyphrases found per document in CiteSeerX.}
\begin{tabular}[t]{c|cccccc}
\hline
Set ID & Min & Q1 & Q2 & Mean & Q3 & Max. \\
\hline \hline
A & 0 & 4 & 7 & 7.409 & 10 & 28 \\
B & 0 & 5 & 8 & 8.313 & 11 & 31 \\
\hline
\end{tabular}
\end{center}
\end{table}

\begin{longtable}[tb]{>{\centering\arraybackslash}m{2cm}||>{\centering\arraybackslash}m{2cm}>{\centering\arraybackslash}m{2cm}>{\centering\arraybackslash}m{2cm}>{\centering\arraybackslash}m{2cm}>{\centering\arraybackslash}m{2cm}}
\caption{\label{tab:expert_example}The top 5 experts of 20 sample queries returned by CSSeer} \\
\hline
Query & 1 & 2 & 3 & 4 & 5 \\
\hline \hline
compiler & Ken Kennedy & S. Amarasinghe & Alok Choudhary & C.-w.\ Tseng & W.-m.\
W.\ Hwu \\
\hline
computer network & K.\ Ramakrishnan & David L.\ Mills & M\'{a}rk Jelasity & Anna Karlin & Karl Levitt \\
\hline
data structure & Martin Rinard & Viktor Kuncak & G.\ St{\o}lting Brodal & Lars
Arge & J.\ Scott Vitter \\
\hline
database & David J.\ Dewitt & Jiawei Han & Serge Abiteboul & L.\ Bertossi & C.\
S.\ Jensen\\
\hline
information retrieval & W.\ Bruce Croft & Jamie Callan & Alan F.\ Smeaton & E.\ Kushilevitz & Yuval Ishai \\
\hline
intelligent agent & Lin Padgham & Michael Winikoff & M.\ Wooldridge & Tim Finin & Milind Tambe \\
\hline
linear algebra & Jack Dongarra & David Walker & James Demmel & R.\ C.\ Whaley & Antoine Petitet \\
\hline
machine learning & Andrew Mccallum & R.\ J.\ Mooney & Peter Stone & R.\ Michalski & Pat Langley \\
\hline
markov chain monte carlo & Jeffrey Rosenthal & Simon J.\ Godsill & G.\ O.\
Roberts & A.\ Doucet & C.\ P.\ Robert \\
\hline
nonparametric statistics & Stefan Schaal & S.\ Vijayakumar & C.\ G.\ Atkeson &
David M.\ Blei & R.\ T.\ Whitaker \\
\hline
programming language & Margaret Burnett & B.\ C.\ Pierce & Frank Pfenning &
Peter Sewell & W.\ Clinger \\
\hline
quality of service & A.\ T.\ Campbell & D.\ C.\ Schmidt & Geoff Coulson & Aurel
Lazar & K.\ Nahrstedt \\
\hline
security & Ran Canetti & D.\ Pointcheval & Gene Tsudik & Mihir Bellare & David Wagner\\
\hline
semantic web & Tim Finin & Steffen Staab & Li Ding & Anupam Joshi & Dieter Fensel \\
\hline
social network & Jennifer Golbeck & Mitsuru Ishizuka & Yutaka Matsuo & Peter A.\ Gloor & David Kempe \\
\hline
software engineering & Victor R.\ Basili & M.\ Wooldridge & N.\ R.\ Jennings &
M.\ Zelkowitz & Reidar Conradi \\
\hline
support vector machine & Glenn Fung & O.\ Mangasarian & Yi Lin & K.\ P.\ Bennett & Grace Wahba \\
\hline
virtual machine & Mendel Rosenblum & Jay Lepreau & Godmar Back & Mike Hibler &
P.\ Tullmann \\
\hline
VLSI & Andrew B.\ Kahng & Jason Cong & Christof Koch & G.\ Indiveri & Igor L.\ Markov\\
\hline
world wide web & Mark Crovella & B.\ Mobasher & Azer Bestavros & Dayne Freitag & Dieter Fensel \\
\hline
\end{longtable}

\begin{longtable}{p{1.2cm}|p{9.6cm}|p{2.3cm}}
\caption{\label{tab:expertise_example}Top 15 expertise list of 10 selected authors} \\
\hline
Author Name & Top-15 Expertise & Note \\
\hline \hline
Ian T.\ Foster & resource management, distributed computing, parallel computer, web service, message passing, distributed system, quality of service, application development, high performance, web services, data management, data transfer, distributed systems, grid computing, high performance fortran & Most cited computer scientist by MAS \\
\hline
Ronald L.\ Rivest & block cipher, public key, encryption key, radio frequency, mobile robot, digital signature, binary relation, secret key, error rate, efficient algorithm, advanced encryption standard, initialization vector, hash function, learning algorithm, probability distribution & 2nd most cited computer scientist by MAS \\
\hline
Scott J.\ Shenker & admission control, congestion control, sensor network, routing algorithm, degree distribution, distributed system, network topology, routing protocol, hash table, wireless sensor network, building block, direct product, denial of service, zipf's law, quality of service & 3rd most cited computer scientist by MAS \\
\hline
Jeffrey D.\ Ullman & information sources, data model, query language, synthetic data, database system, information retrieval, data mining, object model, case study, random sampling, performance analysis, collaborative filtering, efficient algorithm, next generation, association rules & 4th most cited computer scientist by MAS \\
\hline
Jiawei Han & data mining, association rule, association rules, knowledge discovery, data stream, efficient algorithm, clustering algorithm, information system, query processing, data warehousing, time series, data analysis, database system, web page, classification accuracy & Most search person and 3rd highest H-index by ArnetMiner \\
\hline
Pat Langley & machine learning, process model, recommendation system, nearest neighbor, knowledge base, learning algorithm, intelligent system, artificial intelligence, reinforcement learning, mobile robot, domain knowledge, data mining, knowledge discovery, bayesian network, feature selection & 2nd most search person by ArnetMiner \\
\hline
Vladimir Vapnik & support vector, feature selection, time series, radial basis function, prior knowledge, basis function, feature space, learning algorithm, gradient descent, density estimation, pattern recognition, cost function, dna microarray, object recognition, model selection & 3rd most search person by ArnetMiner \\
\hline
W.\ Bruce Croft & information retrieval, natural language, query processing, data management, natural language processing, bayesian inference, topic model, network model, search engine, optical character recognition, machine learning, information system, system performance, dynamic environment, formal model & 4th most search person by ArnetMiner \\
\hline
Anil K.\ Jain & pattern recognition, feature extraction, clustering algorithm, face recognition, feature selection, performance evaluation, image database, computer vision, classification accuracy, error rate, feature vector, similarity measure, statistical learning theory, case study, image analysis & Highest H-index person by ArnetMiner \\
\hline
Hector Garcia-Molina & information sources, data model, information retrieval, digital library, query processing, data warehousing, information system, query language, change detection, search engine, index structure, digital document, electronic commerce, efficient algorithm, web search & 2nd highest H-index person by ArnetMiner \\
\hline
\end{longtable}

The Wikipedia based keyphrase candidates are usually highly meaningful terms
since Wikipedia titles and link texts are manually edited.  However, the
coverage (or how well these terms cover the topics in the given discipline) is
unknown.  Although we intentionally include Wikipedia pages related to Computer
Science, Statistics, and Mathematics, whether these pages are adequate topics to
represent most CiteSeerX documents is still an unanswered question.

In order to answer this question, we begin by studying the distribution of the number of the keyphrases found in a document.  We randomly select $10,000$ documents as Set A from CiteSeerX.  Using only the title and the abstract, we count the number of keyphrases found for each document using only keyphrase candidates compiled from Wikipedia.

Figure~\ref{fig:keyphrase_per_doc_no_prune} demonstrates the empirical
distribution of the number of keyphrases found per document in Set A.  As shown,
less than $4\%$ of documents do not have any matched keyphrases.  Half of the
documents have at least 7 matched keyphrases.  On average, a document has 7.409
matched keyphrases using only the title and the abstract.

To further study the documents with 0 or few keyphrase matches, we randomly sample
100 documents that have no keyphrase matches and examine the contents.  We found that 74 out of the 100 documents are parsed incorrectly in the PDF to text process.
A typical mistake is an extremely short title or abstract, or even empty title
and abstract.  Other cases include missing spaces between words, and contents
with garbage or unreadable characters.  For the rest of the documents, most of them are not valid papers, scanned papers, or papers written in foreign languages.

To study the Wikipedia based keyphrase extraction strategy without the influence
of extremely short titles or abstracts, we compile Set B from $10,000$ randomly
sampled documents whose titles have at least 4 words and abstracts have at least
20 words.  The probability mass function of keyphrases found per document is
shown in Figure~\ref{fig:keyphrase_per_doc_min_length}.  Only $0.4\%$ of sampled
documents have no matched keyphrases.  Half of the documents have at least 8
matched keyphrases, and on average a document has 8.313 keyphrases.  Since the
keyphrase extractor can retrieve a decent number of keyphrases using only the
title and the abstract of a document, Wikipedia is a promising resource for
keyphrase candidate compilation for scientific literature.

The detail of the number of keyphrases found per document in the two sets is
shown in Table~\ref{tab:keyphrase_per_doc}.

\section{Case Study} \label{sec:case}

We illustrate sample outputs of an expert list and an expertise list to show the
practicality of the system.

\subsection{Expert List}

We start the case study by showing several examples of an expert list returned by
CSSeer.  As shown in Table~\ref{tab:expert_example}, 20 different terms ranging
in several different sub-domains of computer science are selected as query
terms.  We report the top 5 returned experts.

To measure whether the returned names are experts of the given query, we
manually checked each of these researchers' homepage and their total number of
citations compiled by MAS.  If the query term appears in the person's homepage
and the author's total number of citations is larger than $500$, it is very
likely that the researcher is a good candidate for an expert of the given area.

From the researchers' homepages, we found only 5 authors whose homepages do not
contain the query term: Stefan Schaal (nonparametric statistics), S.\
Vijayakumar (nonparametric statistics), C.\ G.\ Atkeson (nonparametric
statistics), Christof Koch (VLSI), and Anna Karlin (computer network).  After
carefully examining their profile, 4 of these are actually experts in the query
area, and the synonyms or similar terms of the query appear in their homepage.
The only possible exception is Dr.\ Christof Koch, an expert of Biology and
Engineering.  However, he co-authored a few of highly cited VLSI papers back in
1990s.

As for number of citations, the only two researchers who have less than $500$
citations are Dr.\ Aurel Lazar (4 citations) and Dr.\ K.\ Ramakrishnan (0
citations).  We believe these are MAS's mistakes because at the time of writing,
Dr.\ Lazar has $3,622$ citations and Dr.\ Ramakrishnan has $3,440$ citations by
ArnetMiner.

\subsection{Expertise List}

An expertise list is very helpful for users to learn what an author's research
interest is.  In this section, we show examples of the expertise list of 10
selected authors.  Specifically, from MAS we selected the four most cited computer
scientists (Ian T.\ Foster, Ronald L.\ Rivest, Scott J.\ Shenker, and Jeffrey
D.\ Ullman), from ArnetMiner we selected the top four search people (Jiawei Han, Pat Langley, Vladimir Vapnik, and W.\ Bruce Croft) and three authors who have the
highest H-index (Anil K.\ Jain, Hector Garcia-Molina, and Jiawei Han).  Note that
Dr.\ Jiawei Han is both the 3rd highest H-index author and one of the most searched people by ArnetMiner.  Thus, we ended up collecting 10 names in total for the case study.

We briefly introduce these authors so that readers may examine the extracted top
15 terms and check if they truthfully reflect these authors' expertise.  Dr.\
Foster is famous for the acceleration of discovery in a networked environment
and contributes a lot in high-performance distributed computing, parallel
computing, and grid computing.  Dr.\ Rivest is one of the inventors of the RSA
algorithm and many symmetric key encryption algorithms.  Dr.\ Shenker contributes
much to network research, especially in Internet design and architecture.  Dr.\
Ullman is known for database theory and formal language theory and is an author
of several textbooks in these fields.  Dr.\ Han and Dr.\ Langley are famous for
their contributions in machine learning and data mining fields.  Dr.\ Vapnik
developed the theory of Support Vector Machine.  Dr.\ Croft is well known for
contributions to the theory and practice of information retrieval.  Dr.\ Jain is
a contributor to video encoding, computer vision, and image retrieval.  Dr.\
Garcia-Molina is notable for information management and digital libraries.

The selected authors' top 15 expertise are listed in
Table~\ref{tab:expertise_example}.  As can be seen, the automatically selected
terms on average represent each author's fields of expertise appropriately.  A
user, even without knowing these authors in advance, should be able to tell each
of these authors' research interest by only examining the list of terms.

\section{Conclusions and Future Works} \label{sec:conc}

We describe ExpertSeer, an open source expert recommender system based on
digital libraries.  Using the framework, we built two systems: CSSeer, an expert
recommender for Computer Science, and ChemSeer, an expert recommender for
Chemistry.  The system efficiently handles millions of documents and authors.
We thoroughly investigated CSSeer with the other two state-of-the-art expert
recommender systems, ArnetMiner and Microsoft Academic Search.  We found that
the three systems have moderately diverse opinions on experts for our benchmark
query term set.  This does not mean one system is better or worse than others.
In practice, different expert recommender systems may be biased toward certain
topics or certain authors due to differences in collected data, extraction
methods, ranking, and other analysis.  For a more comprehensive expert list,
users should consider using several systems. Or possibly, a meta-expert list
could be created.  In addition, the related keyphrase list provided by
ExpertSeer could be a promising alternative, since integrating both the experts
of a given query and the experts of the related keyphrases is more likely to
generate a complete expert list.

To quantify the performance of different systems, we compared three recommendation systems and GS* -- a simulating system that imitates Google Scholar's ranking function -- in terms of Precision-at-$k$.  We found that all three real systems reported reasonably good results for top 3, top 5, and top 10 returns, even though the returned name set of each system was moderately different.  Our proposed system has the best performance among these expert recommenders.  The simulating system GS* has a mediocre performance, probably because it does not differentiate in which domains an author has received the citations.  Thus, when an author is outstanding in one area, the authority scores of her other research areas, which are probably less remarkable, will be boosted as well.  Thus, the expert list returned by GS* may include authors who are experts of less relevant fields.

So far, ExpertSeer uses only author-to-document authoring relationship and
document-to-document citation relationship for expert recommendation.  Other
linguistic techniques and heterogeneous social network mining techniques should
also be investigated.  For example, the Bayes' rule can naturally integrate the
reputation of the published conferences or journals into the model.

We cannot access the exact expert ranking functions of ArnetMiner and Microsoft
Academic Search.  Thus, we could only rely on their previous publications to
infer these ranking functions.  In addition, we could only employ their online
services to obtain their recommended expert list.  However, the expert list may
be influenced by several factors besides the ranking function, such as the
collected documents and the author disambiguation algorithm.  Assuming we will
have access to their ranking functions, we can better compare different ranking
functions based on the same document set to eliminate other confounding factors.

Several research questions and applications can be developed based on this
framework.  For example, the influence maximization problem on large-scale
social networks has been widely studied recently~\cite{chen2012information},
\cite{kim2013scalable}.  Since the authors and their expertise lists are identified,
it would be interesting to observe and study how scholars collaborate and influence each other.  In addition, a time factor can be integrated
into the system so that the flow of information from one domain to another
domain can be learned and visualized, and hopefully be used to discover useful
interacting patterns among different research domains.  ExpertSeer can also be
the foundation and provide reliable data source for research in finding
teams of experts in social networks~\cite{lappas2009finding}.

\bibliographystyle{Science}
\bibliography{sigproc}


\clearpage

\end{document}